\documentclass[showpacs,aps,prl,twocolumn,groupedaddress,amsmath,amssymb]{revtex4}
\pdfoutput=1
\usepackage{esint}
\usepackage{graphicx}
\usepackage{soul}
\usepackage{color}
\usepackage{natbib} 
\usepackage[latin1]{inputenc} 
\usepackage{bbm}
\usepackage[unicode=true,
 bookmarks=true,bookmarksnumbered=false,bookmarksopen=false,
 breaklinks=false,pdfborder={0 0 1},backref=false,colorlinks=false]
 {hyperref}

\newcommand{\ev}[1]{\ensuremath{\langle #1 \rangle}}

\newcommand{\PD}{\phantom{\dag}}

\newcommand{\ACcomment}[1]{}

\def    \bse{\begin{subequations}}
\def    \ese{\end{subequations}}

\def \hH{ \hat{\mathcal{H}}}
\def \wM{\omega_{\rm M}}
\def \CC{\mathcal{C}}
\def \tomega{\tilde{\omega}}
\def  \RG{r_{G}}

\begin{document}
\title{Quantum-Limited Amplification via Reservoir Engineering}
	
\author{A. Metelmann}
\email[]{anjam@physics.mcgill.ca}
\author{A. A. Clerk}
\affiliation{Department of Physics, McGill University, 3600 rue University, Montr\'{e}al, Quebec, H3A 2T8 Canada}
\date{\today}

\begin{abstract}
We describe a new kind of phase-preserving quantum amplifier which utilizes dissipative interactions in a parametrically-coupled three-mode bosonic system.  
The use of dissipative interactions provides a fundamental advantage over standard cavity-based parametric amplifiers:  large photon number gains are possible
with quantum-limited added noise, with no limitation on the gain-bandwidth product. 
We show that the scheme is simple enough to be implemented both in optomechanical systems and in superconducting microwave circuits.
\end{abstract}
	

	
\maketitle

	
\textit{Introduction-- }  The past few years have seen a resurgence of interest in amplifiers working near the fundamental limits set by quantum mechanics 
\cite{PhysRevD.26.1817}, in contexts varying from quantum information processing in circuits \cite{Clarke2008,RevModPhys.82.1155,DevoretScience2013}, 
to radio astronomy \cite{Clarke2010}, to ultra-sensitive force detection (e.g., for gravity wave detection \cite{Abadie2011}).  The standard paradigm for a 
quantum-limited, phase-preserving amplifier is the non-degenerate parametric amplifier (NDPA) \cite{PhysRev.124.1646,PhysRev.129.481,PhysRev.160.1076,PhysRev.160.1097}, 
which is based on a coherent interaction involving three bosonic modes (pump, signal and idler).  This interaction simply converts a pump mode photon into two photons, one in the signal mode, the other in the idler mode.  The result is that weak signals incident on the signal mode are amplified, with 
the minimum possible added noise.  There has been remarkable progress in realizing such amplifiers using superconducting circuits \cite{PhysRevLett.60.764,PhysRevA.39.2519,PhysRevLett.65.1419,
Castellanos-Beltran2008,Nakamura2008,Bergeal2010,Bergeal2010a,Hatridge2011}.  This in turn has enabled a number of breakthroughs, from the measurement of mechanical motion near the quantum limit \cite{Teufel2009}, to the measurement of quantum jumps of a superconducting qubit \cite{PhysRevLett.106.110502,Hatridge2013}
 and the implementation of quantum feedback schemes \cite{Vijay2012,PhysRevLett.109.050506}.  

Despite their many advantages, standard cavity-based parametric amplifiers suffer from the limitation of having a fixed gain-bandwidth product:  as one increases the gain of the amplifier, one also reduces the range of signal frequencies over which there is amplification.  This is a fundamental consequence of the amplification mechanism, which involves introducing effective negative damping to the signal mode.  The consequent reduced damping rate determines the amplification, but also sets the amplification bandwidth (see, e.g., \cite{RevModPhys.82.1155}).  This tradeoff between gain and bandwidth can severely limit the utility of cavity-based parametric amplifiers in many applications.
Traveling-wave parametric amplifiers (TWPAs) \cite{Cullen1958,Tien1958} do not use cavities and are in principle not limited in the same way.  In practice however,  good device performance and bandwidth of TWPAs is limited by the requirement of phase-matching (though see Ref.~\onlinecite{Jonas2012} for recent progress in the microwave domain).

 \begin{figure}[t] 
 \centering
 \includegraphics[width=0.99\columnwidth]{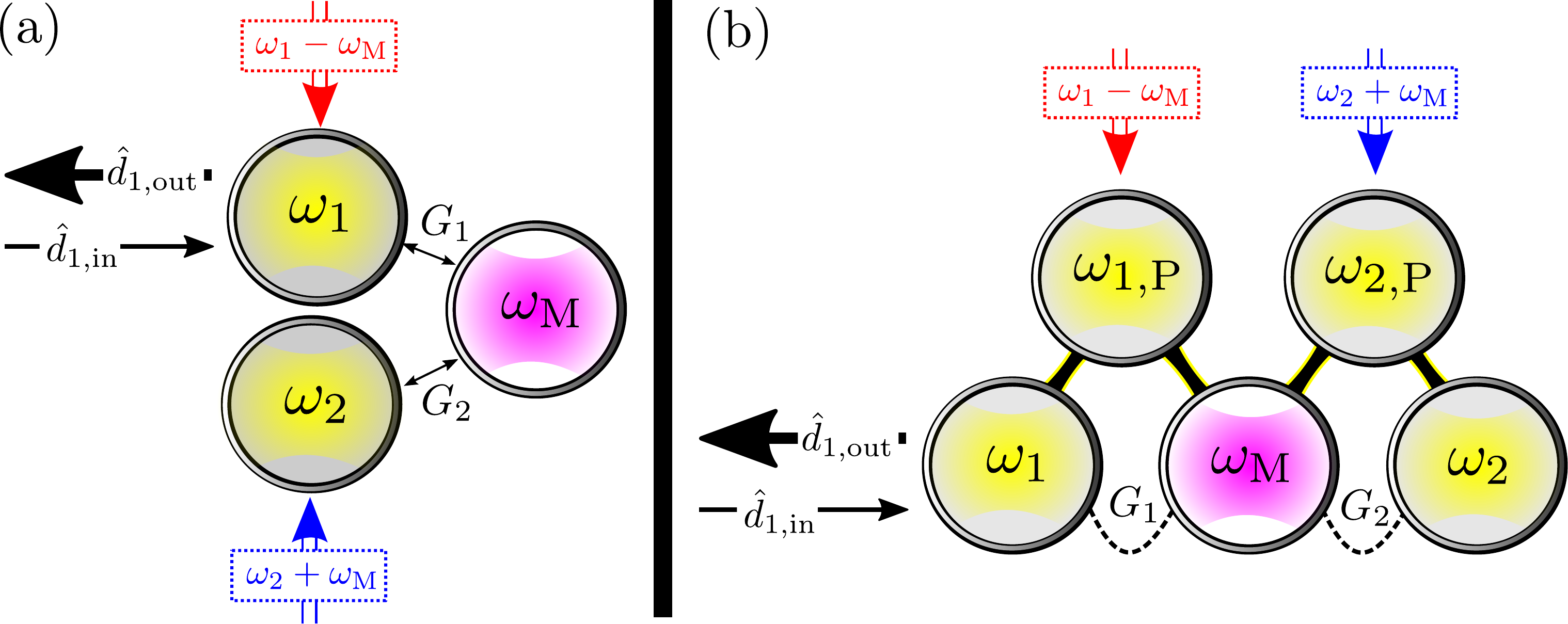}
	\caption{(a) Schematic showing the optomechanical realization of the dissipative amplification scheme.  
	Two driven cavities ($1,2$) are both
	coupled parametrically to a third auxiliary mechanical mode.  The mechanics mediates a dissipative interaction between
	modes $1$ and $2$.  Signals incident on either cavity are amplified in reflection.  
	(b) Alternate realization, where two pump modes $\omega_{1,\rm P}, \omega_{2, \rm P}$  are used to generate the required interaction Hamiltonian, Eq.~(2);
	this setup could be directly implemented using superconducting microwave circuits \cite{EPAPS}.
	\label{Img_1}}
\end{figure}

In this work, we introduce a new approach for quantum-limited amplification based now on three localized bosonic modes.  Unlike a NDPA, our scheme explicitly involves dissipative 
(i.e.~non-Hamiltonian) interactions between the modes.  We show that this approach allows a large gain with quantum limited noise, but crucially  {\it is not limited by a fixed gain-bandwidth product}:  the gain can be arbitrarily large without any corresponding loss of bandwidth. 
Note that non-Hamiltonian evolution is also utilized in a very different way in probabilistic amplifiers \cite{Fiurasek2004,Ralph2009,Xiang2010, Pandey2013}, which can 
stochastically amplify signals without adding noise.

 Our approach is related to reservoir engineering \cite{PhysRevLett.77.4728}, where one constructs a non-trivial dissipative reservoir that relaxes a system to a desired target state 
 (e.g.~an entangled state \cite{Plenio2002,Kraus2004,Parkins2006,PhysRevA.83.052312, PhysRevLett.107.080503}).  
 Here, we instead construct an engineered reservoir which mediates a dissipative amplification process.  Our mechanism can also be interpreted as a kind of coherent feedback process \cite{Lloyd2000,James2008,Mabuchi2008,Nurdin2009}, where the amplification is the result of an autonomous quantum non-demolition (QND) measurement combined with a feedback operation.  Our scheme is simple enough to be realized using existing experimental capabilities, either with three-mode  optomechanical systems (where a mechanical mode couples to two electromagnetic cavity modes) \cite{Painter2012NatComm,Dong2012}, or with superconducting circuits \cite{PhysRevB.87.134504,PhysRevB.79.184301,Bergeal2010}.


\textit{Model-- }While our scheme is amenable to many possible realizations, we focus here for concreteness on a three-mode optomechanical system. 
Two cavity modes (frequencies $\omega_1$ and $\omega_2$), are coupled to a single mechanical mode $\wM$, cf.~Fig.~1.  The cavity photons interact with the mechanical mode via radiation pressure forces, and the system is described by the Hamiltonian $\hH = \hH_{\rm S}  + \hH_{\rm diss}$.  Here, $\hH_{\rm S}$ is the coherent system Hamiltonian ($\hbar=1$),
\begin{align}\label{Eq_1}
\hH_{\rm S} =&   \sum_{j = 1,2} \left\{ \omega_j + g_j \left(\hat b + \hat b^{\dag} \right) \right\} \hat a_j^{\dag} \hat a_j^{\PD} 
			+ \wM  \hat b^{\dag} \hat b,
\end{align}
where $\hat b$ ($\hat a_j$) is the annihilation operator for the mechanical resonator (cavity $j$), and $g_j$ is the optomechanical coupling strength for cavity $j$.  $\hH_{\rm diss}$ describes the damping of all three modes (each by independent baths), and the laser drives on the two cavity modes; these are treated at the level of standard input-output theory \cite{PhysRevA.31.3761,RevModPhys.82.1155}, resulting in cavity (mechanical) damping rates $\kappa_j$ ($\gamma$).

In what follows, the two cavity modes will play roles similar to  ``signal" and ``idler" modes in a NDPA, whereas the mechanics will be used to mediate an effective interaction between them.  To achieve this, we assume a strong coherent drive on each cavity, detuned to the red (blue) mechanical sideband for cavity $1$ ($2$), 
(i.e.~drive frequencies  $\omega_{L,1/2} = \omega_{1/2} \mp \wM)$). We work in an interaction picture with respect to the free Hamiltonians, and
perform displacement transformations:  
$\hat a_j \equiv  \bar a_{j} e^{\pm i\wM t} + \hat d_j $, where $\bar{a}_j$ is the average classical amplitude of cavity $j$ due to the laser drive; we take these to be real without loss of generality.  Assuming the standard experimental situation where $g_j$ are small and $\bar{a}_j$ are large, we linearize $\hH_{\rm S}$, resulting in 
\begin{align} \label{eq:HLinearized}
\hH_{\rm S} =&  \  G_1  \left(  \hat d_1^{\PD}  \hat b^{\dag} +  \hat d_1^{\dag}   \hat b \right) 
                        +  G_2  \left(  \hat d_2^{\PD}  \hat b        +  \hat d_2^{\dag}   \hat b^{\dag} \right) + \hH_{\rm CR}.
\end{align}
Here, $G_j = g_j \bar{a}_j$ are the many-photon optomechanical couplings, and $\hH_{\rm CR}$ describe non-resonant interaction processes.
We focus on the the good-cavity limit $\wM \gg \kappa_j, \gamma$, where the effects of $\hH_{\rm CR}$ will be negligible.  We will thus start by dropping $\hH_{\rm CR}$ for transparency, i.e.~we make the rotating wave approximation (RWA); full results beyond the RWA are presented in the figures and in the EPAPS \cite{EPAPS}.

If $G_1 = 0$, the Hamiltonian of Eq.~(\ref{eq:HLinearized}) describes an optomechanical NDPA, with the mechanics acting as idler; this was recently realized by Massel et al \cite{Sillanpaa2011}.  One might guess that turning on the beam-splitter interaction with cavity 1 by making $G_1 \neq 0$ would simply act to laser cool and optically damp the mechanical mode \cite{Marquardt2007,WilsonRae2007}, 
but not fundamentally change the amplification physics.  This is not the case:  as we show below, the coherence between the control lasers leads to a completely new mechanism.   For $G_1 \geq G_2$, the interactions in Eq.~(\ref{eq:HLinearized}) have been discussed as a means to generate photonic entanglement
 \cite{Parkins2006,Wasilewski2009,Vitali2011,Vitali2012,PhysRevA.87.033829,PhysRevLett.110.253601,Tian2013}; amplification was not discussed.  In contrast, we focus on the case $G_1 = G_2$; while this only leads to minimal intracavity entanglement \cite{PhysRevLett.110.253601}, it is optimal in allowing the mechanics to mediate amplifying interactions between the two cavities.


{\it Dissipative interactions-- } If the mechanical resonator was strongly detuned (in the interaction picture) from the two cavity modes by a frequency $\Delta$, then standard adiabatic elimination of the mechanics would yield the NDPA Hamiltonian, $ \hH^{\rm PA} = \widetilde G \ \hat d_1 \hat d_2 + h.c $, with $\widetilde{G} \sim G^2 / \Delta$.  In contrast, we are interested in the resonant case, where the induced interactions are more subtle.  As the system is linear, one can exactly solve the Heisenberg-Langevin equations corresponding to Eq.~(\ref{eq:HLinearized}), and use these to derive effective equations for the cavity modes with the mechanics eliminated \cite{EPAPS}.  We first consider the simple limit where 
$\gamma \gg \kappa, G$; this results in effectively instantaneous induced interactions.  We also specialize to the ideal case where $\kappa_1 = \kappa_2 \equiv \kappa$ (see \cite{EPAPS} for $\kappa_1 \neq \kappa_2$).  Introducing the effective coupling rate $\Gamma = 4 G^2 / \gamma$, the resulting Langevin equations for the cavity modes are:
\bse
\label{eqs:CavityEOMs}
\begin{align} \label{eq:CavityEOM1}
\dot{\hat d}_1^{\PD} &=  -\frac{( \kappa + \Gamma)}{2}  \hat d_1^{\PD}  - \frac{ \Gamma}{2} \hat d_2^{\dag}
			  - \sqrt{\kappa} \hat d_{1,\rm in}^{\PD} 
			  + i \sqrt{\Gamma} \hat b_{\rm in}^{\PD},
 \\
\dot{\hat d}_2^{\PD} &=  -\frac{( \kappa - \Gamma) }{2} \hat d_2^{\PD} + \frac{ \Gamma}{2} \hat d_1^{\dag}
			  - \sqrt{\kappa} \hat d_{2,\rm in}^{\PD} 
			  + i \sqrt{\Gamma} \hat b_{\rm in}^{\dag}.
\label{eq:CavityEOM2}
\end{align}
\ese
The operators $\hat d_{j,\rm in}$ ($\hat b_{\rm in}$) describe the quantum and thermal noise incident on the two cavities (the mechanics); they have zero mean and correlation functions 
$\ev{\hat o_{\rm in}^{\PD}(t)\hat o_{\rm in}^{\dag}(t')}= \ev{\hat o_{\rm in}^{\dag}(t) \hat o_{\rm in}^{\PD}(t')} + \delta(t-t')  = \delta(t-t') (\bar n_o^T + 1)$, where $o = d_j, b$, and $\bar n_o^T$ is the thermal
occupancy of each bath.

The mechanical resonator gives rise to two effects in Eqs.~(\ref{eqs:CavityEOMs}).  First, it gives rise to an additional positive damping $\Gamma$ of mode 1, and an additional negative damping $-\Gamma$ of mode 2; each effect corresponds simply to one of the two terms in Eq.~(\ref{eq:HLinearized}).  In contrast, the joint action of both interaction terms gives rise to terms in Eqs.~(\ref{eqs:CavityEOMs}) reminiscent of a NDPA, where $\hat{d}_1$ is driven by $\hat{d}^\dagger_2$ and vice versa.  Note crucially the opposite sign of this term in Eq.~(\ref{eq:CavityEOM1}) versus Eq.~(\ref{eq:CavityEOM2}); this difference implies that these terms {\it cannot} be derived from an  NDPA interaction Hamiltonian $\hH^{\rm PA}$.  Instead, they correspond to an effective {\it dissipative} parametric interaction.  Such terms can be obtained from Lindbladian dissipators in a quantum master equation \cite{EPAPS}; they are also sometimes referred to as a phase-conjugating 
interaction \cite{2013arXiv1305.1969B}.  On their own, such terms cause a coherent rotation between $\hat{d}_1$ and $\hat{d}_2^\dagger$, and as such no amplification.  However, 
when combined with the mechanically-induced damping/anti-damping terms, one finds a striking result:  the linear system described by Eqs.~(\ref{eqs:CavityEOMs}) always gives rise to exponential decay in the time domain at a rate $\kappa/2$, {\it irrespective} of the value of $\Gamma$ \cite{EPAPS}.  Thus, unlike a standard paramp, the mechanically-induced cavity-cavity interactions here do not give rise to a slow system decay rate, and do not cause any instability (i.e.~the linear system is stable for all values of $\Gamma$).  This
conclusion holds even when $\gamma / \kappa$ is finite:  the system decay rates are independent of $G$ \cite{EPAPS}.  


{\it Scattering properties-- }While the mechanically-induced interactions do not yield any net anti-damping, they do nonetheless enable amplification.  We use standard input-output theory to calculate the scattering matrix $\mathbf{S}[\omega]$ which relates output and input fields.  For simplicity, we first neglect internal cavity losses.  Introducing the cooperativity $\mathcal C = 4 G^2/(\kappa \gamma)$, and defining the input/output vectors $ \mathbf{\hat D}_{l} \equiv (\hat d_{1,l}^{\PD} ,\hat d_{2,l}^\dagger, \hat b_{l}^{\PD})^{\rm T}$  
($ l \in \{\rm{in,out}\}$), we find in the limit $\gamma \gg \kappa, \omega$:
\begin{eqnarray}
	\label{Eq_5}
 \mathbf{\hat D}_{\rm out}[\omega]  & = & \mathbf{S}[\omega]   \mathbf{\hat D}_{\rm in}[\omega], \\
\mathbf{S}[\omega] & = & 
\left(
\begin{array}{ccc}
 	\frac{2 \CC-1-\tomega ^2}{(1- i \tomega )^2} 	& 	\frac{2 \CC}{(1- i  \tomega )^2} 					& \frac{2 i  \sqrt{\CC}}{1 -  i \tomega } \\
 	\frac{-2 \CC}{(1- i  \tomega )^2} 			& 	-\frac{2 \CC+1+ \tomega ^2}{(1- i \tomega )^2} 		& \frac{-2 i \sqrt{\CC}}{1 -  i  \tomega } \\
 	\frac{2 i \sqrt{\CC}}{1-  i \tomega } 			& \frac{2 i \sqrt{\CC}}{1- i \tomega } 					& -1 \\
\end{array}
\right),
\label{eq:SMatrix}
\end{eqnarray}
where $\tomega = 2 \omega / \kappa$.  Note that at $\omega = 0$, the above result holds for any value of $\gamma $; the full expression of $\mathbf{S}[\omega]$ for arbitrary $\gamma$ is given in the EPAPS \cite{EPAPS}.  For $\mathcal C > 1$, $\mathbf{S}[\omega]$ implies that signals incident on either cavity in a bandwidth $\sim \kappa$ around resonance will be amplified and reflected.  For concreteness, we focus on signals incident on cavity 1 (see EPAPS \cite{EPAPS} for the similar case of signals incident on cavity $2$).  The amplitude gain for such a signal at resonance is simply the reflection coefficient $S_{11}[0] = 2\mathcal C-1 \equiv \sqrt{ \mathcal G_{1}[0]}$.  Clearly, the gain can be made arbitrarily large by increasing $\mathcal{C}$ with no corresponding reduction of bandwidth (which remains $\sim \kappa$).  This is in stark contrast to a standard NDPA, and is a direct consequence of the behavior discussed above:  the mechanically-induced interactions do not induce any net negative damping of 
the system.   

While for simplicity we have focused on the case where the mechanical damping $\gamma$ is large, the same physics holds for an arbitrary $\kappa / \gamma$ ratio.  In the limit of
large $\CC$, the photon number gain is well-approximated as:
\begin{eqnarray}
	\mathcal{G}_1[\omega] \equiv |S_{11}[\omega]|^2 
		\simeq		 \frac{\CC^2}{ \left[1 + (2 \omega / \gamma)^2 \right] \left[1 + (2 \omega / \kappa)^2 \right]^2} 	.	
\end{eqnarray}
The effective bandwidth of the gain interpolates between $\kappa$ for $\gamma / \kappa \gg 1$, and $\gamma$ for $\gamma / \kappa \ll 1$.  Our general conclusions still hold:  the gain can be arbitrarily large by increasing $\CC$, and there is {\it no fundamental limitation on the gain-bandwidth product in this system}.


{\it Added noise-- }  Our scheme can also achieve a quantum-limited added noise.  This follows immediately from the $\mathbf{S}$ matrix in Eq.~(\ref{eq:SMatrix}).  As usual, we define the added number of noise quanta of the amplifier by first calculating the noise spectral density of the amplifier output (i.e.~$\hat{d}_{1,\rm out}[\omega]$).  The contributions to this noise from the mechanical and cavity $2$ input noises constitute the amplifier added noise.  Expressing this as an equivalent amount of incident noise in the signal defines the number of added noise quanta $\bar{n}_{\rm add}[\omega]$; the quantum limit on this quantity in the large-gain limit is $\bar{n}_{\rm add}[\omega] \geq 1/2$ \cite{RevModPhys.82.1155}.   We find at zero frequency:
\begin{eqnarray}\label{Eq_7}
	\bar{n}_{\rm add}[0]  &= &
		  \frac{ \left(\sqrt{ \mathcal{G}_1[0]} + 1\right)^2 }{\mathcal{G}_1[0]}  \left( \frac{1}{2} + \bar{n}^T_{d_2} \right) + 
		 \frac{1 + \sqrt{\mathcal G_1[0]}   }{ \mathcal G_1[0] } \left( 1 + 2 \bar{n}^T_{b} \right) \nonumber \\
		& = &
			\frac{1}{2} + \bar{n}^T_{d_2} + \frac{2 + 2 \bar{n}^T_{d_2}+ 2 \bar{n}^T_{b}}{\sqrt{\mathcal{G}_1[0]}}  + \mathcal{O}\left [ \frac{1}{\mathcal{G}_1[0]}\right].
\end{eqnarray}

Thus, if cavity 2 is driven purely by vacuum noise, then in the large gain limit our amplifier approaches the standard quantum limit on a phase-preserving linear amplifier.  On some level, this is surprising.  The ideal performance of a NDPA can be attributed to the fact that it has only a single additional degree of freedom beyond the signal mode \cite{RevModPhys.82.1155}.  In contrast, our system has two additional degrees of freedom (i.e.~idler mode and mechanical mode); one might have expected that the presence of an extra mode would imply extra noise beyond the quantum limit.  That this is not the case highlights the fact that the mechanical mode acts only as a means to mediate an effective dissipative coupling.

It is also worth stressing that in the large $\mathcal{G}_1$ limit,  the contribution of mechanical thermal noise is suppressed by a factor $1/\sqrt{\mathcal{G}_1[0]}$.  This is in stark contrast to the optomechanical NDPA of Ref.~\onlinecite{Sillanpaa2011}.  In that system, the mechanical mode acts as the idler; as such, quantum-limited performance is only possible if the mechanical resonator is at zero temperature, irrespective of the amplifier gain.



To illustrate the effectiveness of our scheme, we show in Fig.~2 expected results for the gain and added noise for a realization based on a microwave-cavity optomechanical system, similar to 
those in Refs.~\onlinecite{Teufel2011,Teufel2011b}.  While such experiments typically have a small mechanical damping rate $\gamma$ (and, hence, small bandwidth), one could use a third auxiliary mode to both laser cool the mechanical mode and enhance its linewidth \cite{PhysRevLett.110.253601}; we have assumed this situation.  One could also use a GHz-frequency, low-$Q$ mechanical resonator (similar to, e.g., Ref.~\onlinecite{Cleland2010}) 
to achieve bandwidths $\sim 10 - 100 \textrm{ MHz}$ \cite{EPAPS}.

\begin{figure*}[ht]  
 \includegraphics[width=1.0\textwidth]{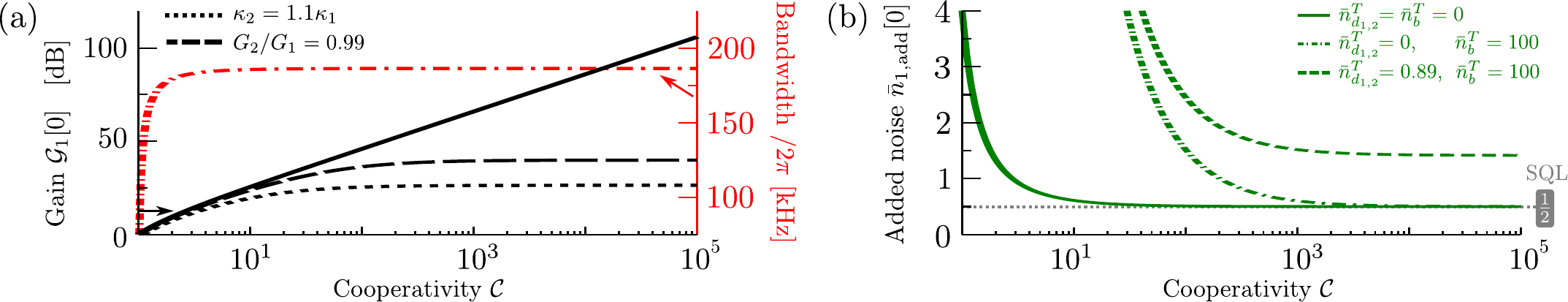}
	\caption{(a) Black curves:  photon number gain versus cooperativity $\mathcal{C}$ for parameters corresponding to a microwave-cavity
	optomechanical realization of the dissipative amplification scheme.  We take $\wM / (2 \pi) = 20 \textrm{ MHz}$ 
	and $\kappa / (2 \pi)  = 1 \textrm{ MHz}$ (solid curve);  the latter
	includes internal loss $\kappa^{\rm int}/( 2 \pi)  = 10 \textrm{ kHz}$.  
	The dotted line includes the effect of asymmetric cavity damping (see legend); the dashed line shows
	instead the effects of $G_1 \neq G_2$ (see legend).  We assume that the mechanical
	resonator is coupled to a third auxiliary cavity which is used to both cool and optically damp it, 
	leading to a total mechanical
	damping rate of $\gamma / (2 \pi) = 200 \textrm{ kHz}$.  
	Red dashed-dotted curve:  bandwidth (defined as the full-width at half-maximum of $\mathcal{G}_1[\omega]$)
	 versus $\mathcal{C}$, same parameters as the solid curve. 
	(b)  Amplifier added noise $\bar{n}_{\rm add}$ versus $\mathcal{C}$.  
	Solid curve:  mechanics and cavity $2$ driven by vacuum noise only.  Dashed-dotted curve:  mechanics now driven by
	thermal noise.  
	Dashed curve:  both mechanics and cavity $2$ driven by thermal noise. 
	Other parameters identical to solid black curves in (a) and with $\bar n^T_{b,d_{1,2}}$ as denoted in the graph.
	All curves are produced without making the RWA (though they are well described by the RWA theory).
		\label{Img_2}}
\end{figure*}

\textit{Connection to QND measurement--}
To provide further intuition on the mechanism underlying our scheme, 
it is useful to consider the dynamics in terms of canonically-conjugate quadrature operators.  We introduce these operators in our interaction picture in the standard way:  $\hat d_j \equiv (\hat X_j + i \hat P_j)/\sqrt{2} $ and 
$\hat b = (\hat{U} + i \hat{V})/\sqrt{2}$.  The interaction Hamiltonian in Eq.~(\ref{eq:HLinearized}) (with $G_1 = G_2 = G$) then becomes
\begin{align}\label{eq:HIntQuads}
	\hH_{\rm int} &= \sqrt{2} G \left( \hat U \hat X_{+} + \hat V \hat P_{-}\right),
\end{align}
where we have introduced joint cavity quadrature operators $\hat{X}_{\pm} = (\hat{X}_1 \pm \hat{X}_2) / \sqrt{2}$, $\hat{P}_{\pm} = (\hat{P}_1 \pm \hat{P}_2) / \sqrt{2}$.

Equation~(\ref{eq:HIntQuads}) lets us understand the importance of having $G_1 = G_2$:  for this choice, $\hat{X}_+$ and $\hat{P}_-$ are QND observables.  They commute with the Hamiltonian and are thus conserved quantities.  The QND interaction allows the mechanical resonator to ``measure" both of these joint cavity quadratures:  the
$\hat{V}$ ($\hat{U}$) quadrature of the mechanical output field will contain information on  $\hat{X}_+$ ($\hat{P}_-$).  

A QND measurement on its own will not generate amplification.  The interaction in Eq.~(\ref{eq:HIntQuads}) does more:  it also performs a kind of coherent feedback operation, where the results of the ``measurement" are used to  displace the unmeasured quadratures $\hat{X}_-$ and $\hat{P}_{+}$.
For example, via the first term in Eq.~(\ref{eq:HIntQuads}), the mechanical 
$\hat{V}$ quadrature measures $\hat{X}_+$: at zero frequency (and ignoring noise), the Heisenberg equations of motion (EOM) yield 
$\hat{V} = -(2 \sqrt{2} G / \gamma) \hat{X}_+$.  
But via the second term in Eq.~(\ref{eq:HIntQuads}), we see that $\hat{V}$ is a force on the $\hat{X}_{-}$ quadrature.  Again, the EOMs at zero frequency yield 
$\hat{X}_- = (2 \sqrt{2} G / \kappa) \hat{V} = - 2 \mathcal{C} \hat{X}_+$.  This directly translates into the ($\omega = 0$) input-output relations
\bse
\begin{eqnarray}
	\hat{X}_{+,\rm out} & = & - \hat{X}_{+,\rm in}, \\
	\hat{X}_{-,\rm out} & = & 4 \mathcal{C} \hat{X}_{+,\rm in} - \hat{X}_{-,\rm in}, 
\end{eqnarray}
\ese
where we neglect mechanical noise contributions.  Thus, the joint measurement plus feedback operation has made $\hat{X}_-$ an amplified copy of $\hat{X}_+$, while leaving the QND observable
$\hat{X}_+$ unperturbed.  In an analogous fashion,
$\hat{P}_+$ becomes an amplified copy of $\hat{P}_-$.  If we now express $\hat{d}_{1,{\rm out}}$ in terms of joint quadratures, we can immediately understand how we obtain amplification. For large $\mathcal{C}$, we have:
\begin{eqnarray}
	\hat{d}_{1,{\rm out}} & = & \frac{1}{2} \sum_{\sigma = \pm} \left( \hat{X}_{\sigma} + i \hat{P}_{\sigma} \right)_{\rm out} 
		\simeq \frac{1}{2} \ \left( \hat X_- + i \hat P_{+} \right)_{\rm out}
	\nonumber \\
	& \simeq &
		\frac{1}{2} \left(4 \mathcal{C} \right) \left( \hat X_+ + i \hat P_{-} \right)_{\rm in} 
		=  2 \mathcal{C} \left( \hat d_1^{\PD} + \hat d_2^{\dag}\right)_{\rm in}.
\end{eqnarray}
Thus, the QND measurement-plus-feedback operations on the joint quadrature operators directly let us understand the structure of the scattering matrix, and the observed amplification.
Note that somewhat analogous QND interactions play a crucial role in the construction of continuous variable cluster states \cite{Zhang2006}.

The QND form of Eq.(\ref{eq:HIntQuads}) also explains the absence of any induced damping of the cavities by the mechanics, see \cite{EPAPS}.
When $G_1 \neq G_2$, the QND nature of the interaction is lost (i.e.~$X_+$, $P_-$ are no longer conserved), 
and thus for fixed $G_2/G_1$, $\mathcal{G}_1[0]$ saturates as a function of $\mathcal{C}_1$.  The same is true when 
$\kappa_1 \neq \kappa_2$.  One finds that in this case, the gain $\mathcal{G}_1[0]$ saturates 
at a value $ [(\kappa_1+\kappa_2) / (\kappa_2 - \kappa_1)]^2$ in the large $\CC$ limit \cite{EPAPS}.
We stress that even with small coupling or damping rate asymmetries, one can achieve very large gains [see Fig.~2(a)] with no loss of bandwidth.  
One can even significantly increase  the amplification bandwidth over the symmetric case, 
yielding amplitude-gain bandwidth products which far exceed $\kappa$ (see EPAPS \cite{EPAPS}).


\textit{Superconducting circuit realization-- }Our scheme could also be realized in a superconducting circuit, where the required interactions 
in Eq.~(\ref{eq:HLinearized}) are realized using Josephson junctions.  Here, the role of the mechanical mode would now also be played by a microwave cavity mode, allowing $\gamma$ to be large.  Further details on such realizations are presented in the EPAPS \cite{EPAPS},
where we show that they offer advantages over conventional Josephson paramps, such as Ref.~\onlinecite{Mutus2013}.  
Using similar parameters to that work, our scheme can achieve quantum-limited amplification with 
a bandwidth of $\sim 47 \textrm{ MHz}$ and a amplitude gain - bandwidth product of 
$\sim 1900 \textrm{ MHz}$, a factor of $3.8$ larger than the device reported in Ref.~\onlinecite{Mutus2013}; unlike Ref.~\onlinecite{Mutus2013}, this performance does 
not require a low-$Q$ signal cavity.  
An optomechanical system using a high-frequency, low-$Q$ mechanical resonator (like in the experiment of Ref.~\onlinecite{Cleland2010}) could also attain similar performance.


\textit{Conclusion-- }
We have described a new method for quantum-limited phase-preserving amplification which utilizes dissipative interactions; unlike standard cavity-based parametric
amplifiers, it does not suffer from any fundamental limitation on the gain-bandwidth product.  
The scheme can be implemented both with optomechanics and with superconducting circuits.


This work was supported by the DARPA ORCHID program through a grant from AFOSR.


\clearpage

\global\long\def\theequation{S.\arabic{equation}}

\global\long\def\thefigure{S.\arabic{figure}}

\setcounter{equation}{0}

\setcounter{figure}{0}

\thispagestyle{empty}
\onecolumngrid
\begin{widetext} 
 \begin{center}
{\fontsize{12}{12}\selectfont
\textbf{Supplemental \hspace{-0.2cm} Material for
	 ``Quantum-Limited \hspace{-0.2cm} Amplification \hspace{-0.2cm} via \hspace{-0.2cm} Reservoir \hspace{-0.2cm} Engineering"\\
[5mm]}}
{\normalsize A. Metelmann and A. A. Clerk\\[1mm]} 
{\fontsize{9}{9}\selectfont  
\textit{Department of Physics, McGill University, Montr\'{e}al, Quebec, Canada H3A 2T8 }} 
 \end{center}   \normalsize  

\section{Gain and noise on the level of linear Langevin equations}
Starting from the system Hamiltonian in RWA approximation, cf. Eq.~(2) without $\hH_{\rm CR}$, and by using input-output theory \cite{PhysRevA.31.3761SI}  to include the dissipative environment, we  derive the quantum Langevin equations for the system. We define the mode operator $ \mathbf{\hat D} \equiv (\hat d_{1}^{\PD} ,\hat d_{2}^{\dag}, \hat b)^{\rm T}$
and include the intrinsic losses $\mathbf{\hat D}_{\xi} = [\hat \xi_{1}^{\rm int},\hat \xi_{2}^{\rm int \dag},0]^{\rm T}$  with the loss rates $\kappa_{j}^{\rm int} (j\in1,2)$ for both cavity modes, as well as the input fluctuations $\mathbf{\hat D}_{\rm in} = [\hat d_{1,\rm in}^{\PD},\hat d_{2,\rm in}^{\dag},\hat b_{\rm in}^{\PD}]^{\rm T}$ with the rates $\kappa^{\rm ex}_{j}$ for the cavities and $\gamma$ for the mechanical system. Here $\kappa^{\rm ex}_{j}$ is associated with the coupling of each cavity to the waveguide used to drive it and to extract signals.
Hence, the total decay rates for the photonic system become $\kappa_{j} = \kappa_{j}^{\rm ex} + \kappa_{j}^{\rm int}$. 

All fluctuations are correlated to thermal baths as denoted in the main text.
As usual, we work in frequency space and end up with the quantum Langevin equations
\begin{align}\label{SI_Eq_1}
  \mathbf{\hat D}[\omega] &=
 \boldsymbol{\widetilde \chi}[\omega]
\left[
\left(
\begin{array}{ccc}
	-\sqrt{\kappa_{1}^{\rm ex} }  & 0                    &   0
\\
	0                             &-\sqrt{\kappa_{2}^{\rm ex} }    & 0
\\
	0                 & 0                   & -\sqrt{\gamma }  \\
\end{array}
\right)
\mathbf{\hat D}_{\rm in}[\omega]
+
\left(
\begin{array}{ccc}
	-\sqrt{\kappa_{1}^{\rm int} }  & 0                    &   0
\\
	0                             &-\sqrt{\kappa_{2}^{\rm int} }    & 0
\\
	0                 & 0                   &0  \\
\end{array}
\right)
\mathbf{\hat D}_{\xi}[\omega]
\right],
\end{align}
introducing the susceptibility matrix 
\begin{align}\label{SI_Eq_2}
 \boldsymbol{\widetilde \chi}[\omega] &= 
\left(
\begin{array}{ccc}
	\chi_1[\omega]^{-1} & 0                    &   i G_1
\\
	0                   &\chi_2[\omega]^{-1}   & - i G_2
\\
	i G_1                 & iG_2                   & \chi_{M}[\omega]^{-1} \\
\end{array}
\right)^{\displaystyle -1} ,
\end{align}
containing the free susceptibilities $\chi_{j}[\omega] = \left[-i \omega + \frac{\kappa_{j}}{2} \right]^{-1}$ and  $\chi_{M}[\omega] = \left[-i \omega + \frac{\gamma}{2} \right]^{-1}$ for the cavities and the mechanical resonator. In this section, we concentrate on a symmetric parameter setting, which means that we assume $\kappa_1 = \kappa_2 \equiv \kappa$ as well as $G_1 = G_2 \equiv G$. Following from the assumption of equal decay rates, the free susceptibilities for both cavity modes coincide, i.e.,~$\chi_1[\omega] = \chi_2[\omega] \equiv \chi[\omega]$,
and the three eigenvalues of the susceptibility matrix, calculated for zero frequency, are 
\begin{align}\label{SI_Eq_3}
 \epsilon_{1,2} = 2/\kappa, \hspace{1cm} \epsilon_3 = 2/\gamma. 
\end{align}
Hence, they are independent of $G$ and the time-dependent solutions are damped and will oscillate as 
$  \mathbf{D}(t) = \mathbf{E}_1 e^{-\kappa/2t} + \mathbf{E}_2 e^{-\kappa/2t} + \mathbf{E}_3 e^{-\gamma/2 t}$.
From this we see that the system is stable irrespective of $G$, because no antidamping is present, 
i.e.,~the eigenvalues of $\boldsymbol{\widetilde \chi}[\omega]$ are always real and positive.

Moreover, we want to calculate the scattering matrix of the system; to keep the resulting expressions compact we neglect intrinsic losses for this derivation. We start from the input-output relations $\hat d_{j,\rm out} = \hat d_{j,\rm in} + \sqrt{\kappa} \hat d_j $ and $\hat b_{\rm out} = \hat b_{\rm in} + \sqrt{\gamma} \hat b $, which connect the  input signals to the respective output signals. Hence, the scattering matrix reads
\begin{equation}\label{SI_Eq_4}
 \mathbf{S}[\omega]=
\left(
\begin{array}{ccc}
	  \displaystyle  \frac{ 2 \mathcal C - (1 - i \frac{2\omega}{\gamma} ) \left( 1 + \frac{4 \omega ^2}{\kappa^2}\right)}
                              { (1 - i \frac{2\omega}{\gamma} ) (1 - i \frac{2\omega}{\kappa})^2}
	& \displaystyle  \frac{2 \mathcal C }{ (1 - i \frac{2 \omega}{\gamma} ) (1 - i \frac{2  \omega}{\kappa} )^2}
	& \displaystyle  \frac{2 i \sqrt{\mathcal C}}{   (1 - i \frac{2 \omega}{\gamma} ) (1 - i \frac{2  \omega}{\kappa} )}
\\[0.2cm]
	  \displaystyle -\frac{2 \mathcal C }{ (1 - i \frac{2 \omega}{\gamma} ) (1 - i \frac{2  \omega}{\kappa} )^2} 
	& \displaystyle -\frac{2 \mathcal C +  (1 - i \frac{2\omega}{\gamma} ) \left(1+ \frac{ 4 \omega^2}{\kappa^2}\right)}
                              {(1 - i \frac{2\omega}{\gamma} ) ( 1 - i \frac{2 \omega}{\kappa} )^2}
	& \displaystyle -\frac{2 i \sqrt{\mathcal C}}{   (1 - i \frac{2 \omega}{\gamma} ) (1 - i \frac{2  \omega}{\kappa} )} 
\\[0.2cm]
	  \displaystyle  \frac{2 i \sqrt{\mathcal C}}{   (1 - i \frac{2 \omega}{\gamma} ) (1 - i \frac{2  \omega}{\kappa} )}
	& \displaystyle  \frac{2 i \sqrt{\mathcal C}}{   (1 - i \frac{2 \omega}{\gamma} ) (1 - i \frac{2  \omega}{\kappa} )} 
	& \displaystyle  1-\frac{2  }{1 - i \frac{ 2 \omega}{\gamma} } \\
\end{array}
\right),
\end{equation}
where $ \mathbf{\hat D}_{\rm out}[\omega] = \mathbf{S}[\omega]  \mathbf{\hat D}_{\rm in}[\omega]$ and 
from which we derived the result in Eq.~(5) of the main text for the limit of  $\gamma \gg \kappa ,\omega$. 
The first (second) row of the scattering matrix equals the output signal of cavity $1$ ($2$). 
The diagonal elements coincide with the reflection coefficient and the off-diagonal terms correspond to the added noise by the amplifier.

Now we include again intrinsic losses, i.e.,~we have then $\kappa = \kappa^{\rm ex} + \kappa^{\rm int}$
and the slightly modified input-output relations  $\hat d_{j,\rm out} = \hat d_{j,\rm in} + \sqrt{\kappa^{\rm ex}} \hat d_j $ for the cavity modes.
Additionally, we want to discuss if differences arise if the input signal is applied either to cavity $1$ or cavity $2$.
The respective gain for both cases yields
\begin{align}\label{SI_Eq_5}
 S_{jj} [\omega ] =& 1 - 2 \frac{\kappa^{\rm ex}}{\kappa}
		      \left[\frac{1}{ 1 - i \frac{2 \omega}{\kappa}} 
			    \mp \frac{\CC}{ \left(1 - i \frac{2 \omega}{\gamma} \right) \left(1 - i \frac{2 \omega}{\kappa} \right)^2} 	\right]
			    \hspace{0.5cm} \Rightarrow \hspace{0.5cm}
\widetilde{ \mathcal G}_j [0] = \left| S_{jj} [0] \right|^2=  \left[1 + \frac{\kappa^{\rm ex}}{\kappa} \left( \sqrt{\mathcal G_j[0]} -1\right)\right]^2	,		    
\end{align}
where the tilded $\widetilde{ \mathcal G}_j [0]$ denotes the gain for $\kappa^{\rm int} \neq 0$ and $\sqrt{\mathcal G_j[0]} = \pm( 2 \mathcal C \mp 1) $ equals the photon number gain without intrinsic losses on resonance. The bandwidth of an output signal either at cavity $2$ or cavity $1$ is equal and, in the regime of large gain, there are no significant differences between the maximal obtained gain.

In the range of intermediate amplification (i.e.,~$\CC$ not much larger than one) the gain maxima differ slightly:
\begin{align}\label{SI_Eq_6}
\widetilde{ \mathcal G}_1[0] - \widetilde{ \mathcal G}_2[0] =  - \frac{8 \kappa^{\rm ex} }{\kappa} \  \left(\frac{ \kappa^{\rm ex} -  \kappa^{\rm int}}{\kappa^{\rm ex} +  \kappa^{\rm int}} \right)  \mathcal C.
\end{align}
Thus, for $\kappa^{\rm ex} > \kappa^{\rm int}$ the amplification gain is slightly larger if the signal is applied to mode $2$.
Keep in mind however that the gain scales with $\mathcal C^2$ and hence the difference in the amount of gain between the two modes is negligible in the limit of large amplification.
To amplify an input signal incident on cavity $1$ one needs $\mathcal C >1$; this condition is independent of the amount of intrinsic losses.
In contrast, amplification of a signal incident on cavity $2$ is slightly more sensitive to intrinsic losses and it exhibits a different threshold for the cooperativity and the decay rates. Amplification of signals incident on cavity $2$ only occurs if $\mathcal C > \kappa^{\rm int}/\kappa^{\rm ex}$, which is a weak condition as one generally wants the coupling $\kappa$ to dominate, 
i.e.,~ $\kappa^{\rm int} \ll \kappa^{\rm ex} $. 

We find a  similar behavior for the corresponding output noise, which we calculate by solving at first the system of Langevin equations in Eq.~(\ref{SI_Eq_1}) and afterwards, we use the input-output relation to obtain the output operators $\hat d_{j,\rm out}$. The corresponding noise spectra are then obtained from
\begin{align}\label{SI_Eq_7}
 \bar S_{j,\rm out}[\omega]      = \frac{1}{2} \ \int \frac{d\Omega}{2\pi} \  \langle \{\hat d_{j,\rm out}^{\PD}(\omega), \hat d_{j,\rm out}^{\dag}(\Omega)  \} \rangle.
\end{align}
To evaluate this, we need the definitions of the noise correlation functions 
$\ev{\hat o_{\rm in}^{\PD}(\omega)\hat o_{\rm in}^{\dag}(\Omega)}= \ev{\hat o_{\rm in}^{\dag}(\omega) \hat o_{\rm in}^{\PD}(\Omega)} + 2 \pi \delta(\omega + \Omega)  = 2 \pi \delta(\omega + \Omega) (\bar n_o^T + 1)$, where $o = d_j, b$. Finally, for an input signal incident on cavity $j$ the corresponding output noise on resonance becomes
\begin{align} \label{SI_Eq_8}
 \bar  S_{j,\rm out} [0] =&   \widetilde{ \mathcal G}_j [0] \left( \bar n^{\rm ex}_{d_j} \hspace{-0.05cm} + \hspace{-0.05cm} \frac{1}{2}\right)
                        \hspace{-0.05cm} + \hspace{-0.05cm}   \frac{\kappa^{\rm int} \kappa^{\rm ex}}{\kappa^2}  \left[ \sqrt{\mathcal G_j[0]} - 1\right]^2 \hspace{-0.1cm} 
                          \left( \bar n^{\rm int}_{d_j} \hspace{-0.05cm} + \hspace{-0.05cm}  \frac{1}{2}\right)
                        \hspace{-0.05cm} + \hspace{-0.05cm}  \frac{ \kappa^{\rm ex}}{\kappa}   \left[ \sqrt{\mathcal G_j[0]} + 1 \right]^2 \hspace{-0.1cm} 
                           \left(\bar n^{T}_{d_{\bar j}} \hspace{-0.05cm} + \hspace{-0.05cm}  \frac{1}{2} \right)
		         \hspace{-0.05cm} + \hspace{-0.05cm}  \frac{2 \kappa^{\rm ex}}{\kappa} \left| \sqrt{\mathcal G_j[0]} + 1 \right|  \hspace{-0.05cm} 
		             \left( \bar n_b^{T} \hspace{-0.05cm} + \hspace{-0.05cm}  \frac{1}{2} \right),
\end{align}
with the definition $\bar n_{d_j}^T = (\kappa^{\rm ex} \bar n^{\rm ex}_{j} + \kappa^{\rm int} \bar n^{\rm int}_j )/\kappa , j \in 1,2$. 
Here, cavity $j$ equals the signal mode and cavity $\bar j$ can be understood as the corresponding idler mode, while the mechanical oscillator is just an auxiliary mode.
The first term in Eq.~(\ref{SI_Eq_8}) is simply the amplified noise of the signal incident on cavity $j$ and the second term originates from the losses inside of this cavity, a contribution which also appears in a standard paramp setup.
The remaining terms in Eq.~(\ref{SI_Eq_8}) are important. They correspond to the added noise by the idler cavity mode as well as the mechanical mode. 
The noise arising due to the latter auxiliary mode scales only linear with the cooperativity, cf. $\sqrt{\mathcal G_j[0]} \sim \CC$, and hence is much smaller than the other contributions. Thus, for a large cooperativity the added noise of the amplifier is quantum limited, as discussed in the main text.

The output noise emanating from cavity $1$ or $2$ differ slightly from each other in the regime of intermediate amplification; similar to the gain maxima, cf. Eq.~(\ref{SI_Eq_6}), the deviations scale linearly with the cooperativity $\mathcal C$. The more relevant question is whether the number of added noise quanta is different for both cavities. Therefore, we  consider the added noise quanta (without intrinsic losses) 
\begin{align}\label{SI_Eq_9}
\bar n_{j,\rm add} \equiv \frac{ \bar  S_{j,\rm out} [0]}{\mathcal G_j [0]} - \left( \bar n^{T}_{d_j} \hspace{-0.05cm} + \hspace{-0.05cm} \frac{1}{2}\right) =   
		       \frac{ 4 \CC^2 }
			    {(2\CC \mp 1)^2}
			    \left(\bar n^{T}_{d_{\bar j}} \hspace{-0.05cm} + \hspace{-0.05cm}  \frac{1}{2} \right)
		      + \frac{ 4 \CC }
			    {(2\CC \mp 1)^2 }
			    \left( \bar n_b^{T} \hspace{-0.05cm} + \hspace{-0.05cm}  \frac{1}{2} \right),
\end{align}
 which sets the arising noise in relation to the resulting gain and the signal's input noise. In the above equation, the $-$(+) sign refers to $j= 1(2)$. 
 The mechanic's and the idler's noise contribution to the added noise are the same whether one uses cavity $1$ or cavity $2$ as the signal mode, but the difference in the respective gain leads to
\begin{align}\label{SI_Eq_10}
 \bar n_{1,\rm add} -  \bar n_{2,\rm add} = \frac{16 \CC^2 }{(4\CC^2 -1)^2} 
                                            \left[ \CC \left( 2 \bar n^{T}_{d} \hspace{-0.05cm} + \hspace{-0.05cm} 1 \right)
                                             + 2 \bar n_b^{T} \hspace{-0.05cm} + \hspace{-0.05cm}  1 \right]
                                          = \frac{\left( 2 \bar n^{T}_{d} \hspace{-0.05cm} + \hspace{-0.05cm} 1 \right)}{\CC} 
                                             + \frac{\left( 2 \bar n^{T}_{b} \hspace{-0.05cm} + \hspace{-0.05cm} 1 \right)}{\CC^2} 
                                             + \mathcal O \left[\frac{1}{\CC^3} \right],
\end{align}
where we assumed $\bar n_{d_1}^T =  \bar n_{d_2}^T  \equiv \bar n_{d}^T $ for the cavity baths. Therewith we see, that in the limit of a large cooperativity the difference between the added noise quanta is negligible. 

Note, we are interested in the regime of large gain and additionally, we want to have $\kappa^{\rm ex} \gg \kappa^{\rm int}$. The influence of intrinsic losses on the gain and noise properties is then not significant. Hence, we neglect them in our further discussions (though they are easily to include as described in this section).


\section{Adiabatically elimination of the mechanical mode and master equation}
For the case that the damping $\gamma$ is large, we can perform an adiabatic elimination of the mechanical mode. We start again from the system Hamiltonian in RWA, 
i.e.,~Eq.~(2) of the main text without $\hH_{\rm CR}$, and calculate the quantum Langevin equations in time-space. Afterwards, we
derive the stationary solution for the mechanical operator (cf. third row in Eq.~(\ref{SI_Eq_1}) for $\omega=0$ and $G_1=G_2 \equiv G$) and obtain
\begin{align}\label{SI_Eq_11}
\hat b \simeq -\frac{2iG}{\gamma} (\hat d_1^{\PD} + \hat d_2^{\dag}) - \frac{2}{\sqrt{\gamma}} \hat b_{\rm in}. 
\end{align}
Inserting this into the Langevin equations for the cavity operators in time-space, we obtain  Eqs.~(3) of the main text.

In the large $\gamma$ regime our system can as well be described by a master equation approach. 
After the elimination of the mechanical mode we obtain a Markovian master equation, which has standard Lindblad form. It contains the cavity decay terms as well as the dissipative parametric amplification contribution:
\begin{align}\label{SI_Eq_12}
 \frac{d}{dt} \hat \rho &= \frac{1}{i \hbar} \left[\hH_{\rm eff}, \hat \rho \right]
                         +  \Gamma  \mathcal L[\hat d_1^{\PD} + \hat d_2^{\dag}] \hat{\rho} + \kappa  \mathcal L[\hat d_1^{\PD}] \hat{\rho} + \kappa \mathcal L[\hat d_2] \hat{\rho},
\end{align}
with the Lindblad super-operator 
$\mathcal L [\hat o ] \hat{\rho} = \hat o  \hat \rho  \hat o^{\dag} - \frac{1}{2} \hat o^{\dag} \hat o \hat \rho - \frac{1}{2} \hat \rho \hat o^{\dag} \hat o  $ and $ \Gamma = 4  G^2 / \gamma$, while  $\hH_{\rm eff} $
just describes the driving of cavity $1$ (or $2$) by an input signal.  Expanding the first Lindblad superoperator in Eq.~(\ref{SI_Eq_12}), the terms involving a single cavity operator yield the
$\Gamma$-dependent damping and antidamping terms in Eqs.~(3) of the main text.  In contrast, the terms involving both cavity 1 and 2 operators give the phase-conjugated interaction between the two cavities.

The above master equation can as well be expressed in terms of joint quadrature operators, which we defined in the main text (see text below Eq.~(8) of the main paper).
We find:
\begin{align}\label{SI_Eq_13}
\mathcal L[\hat d_1^{\PD} + \hat d_2^{\dag}] \hat{\rho}=  \ \mathcal L[\hat X_{+} + i \hat P_-] \hat{\rho}
                                             = \mathcal L[\hat X_{+} ]  \hat{\rho} + \mathcal L[\hat P_{-}] \hat{\rho} + i \left\{ \hat P_{-} \hat \rho \hat X_{+}  - \hat X_{+} \hat \rho \hat P_{-} \right\}.
\end{align}
Here,  $\mathcal L [\hat X_{+} ]$ and $\mathcal L [\hat P_{-} ]$ are standard measurement superoperators:  they describe the measurement of the $\hat X_{+}$ and $\hat P_{-}$ quadratures by the mechanics at the rate $\Gamma$. The remaining terms give rise to the effective feedback dynamics described in the main text.


\section{Unequal decay rates}
\noindent
As discussed in the main text, if $\kappa_1 \neq \kappa_2$, the QND nature of our system is lost  (i.e.,~$\hat X_+$ and $\hat P_-$ are not conserved quantities), as can easily be seen by writing the Langevin equations in terms of quadratures   
\begin{align}\label{SI_Eq_14}
\frac{d}{dt} \hat X_- =& -\frac{\kappa_1+\kappa_2}{4} \hat X_-  - \frac{\kappa_1 - \kappa_2}{4} \hat X_+  - \Gamma \hat X_+,
\hspace{0.5cm}
\frac{d}{dt} \hat X_+ = -\frac{\kappa_1+\kappa_2}{4} \hat X_+  -\frac{\kappa_1-\kappa_2}{4} \hat X_-  ,
\nonumber \\ 
\frac{d}{dt} \hat P_+ =& -\frac{\kappa_1 + \kappa_2}{4} \hat P_+  - \frac{\kappa_1-\kappa_2}{4} \hat P_-  - \Gamma \hat P_-,
\hspace{0.75cm}
\frac{d}{dt} \hat P_- = -\frac{\kappa_1 + \kappa_2}{4} \hat P_-  - \frac{\kappa_1 - \kappa_2}{4} \hat P_+  ,
\end{align}
where we ignored noise terms. The joint quadratures $\hat X_+$ and $\hat P_-$ do not directly couple to the mechanical mode, but due to the mixing of the joint quadrature operators for unequal decay rates the QND nature gets lost and $\hat X_+$ and $\hat P_-$ are not longer conserved quantities.  For $\kappa_1 = \kappa_2$ the quadratures $\hat X_+$ and $\hat P_-$ decouple completely from the system and the QND measurement is restored.

We like to briefly discuss the modifications of the gain and the noise in the case of unequal decay rates. The corresponding expressions are derived in the simplest manner by starting again at the level of Langevin equations in Eq.~(\ref{SI_Eq_1}).
For concreteness, we also focus on the case where the signal is incident on cavity $1$ (similar results follow for the opposite case).  We find for the photon number gain:
\begin{align}\label{SI_Eq_15}
S_{11}[\omega] =      \frac{\left(- i \frac{2\omega}{\kappa} - 1 \right) \left(-i\frac{2\omega}{\kappa} +1+ \frac{\delta \kappa}{\kappa}\right) \left(-i\frac{2\omega}{\gamma} + 1 \right)
                           +  \CC ( 2  + \frac{\delta \kappa}{\kappa} ) }
			    {\left(-i\frac{2\omega}{\kappa} + 1 \right)
			      \left(-i\frac{2\omega}{\kappa} + 1 + \frac{\delta \kappa}{\kappa} \right) \left(-i\frac{2\omega}{\gamma} + 1 \right) 
			   + \CC \frac{\delta \kappa}{\kappa} }
\hspace{0.3cm} \Rightarrow \hspace{0.3cm}
  \mathcal G_1[0] =  \left[\frac{ 2 \CC -1  + \frac{\delta \kappa}{\kappa} [\CC -1] } {   \frac{\delta \kappa}{\kappa}  [\CC  + 1] + 1 }\right]^2,
\end{align}
with $\kappa_2 = \kappa_1 + \delta\kappa \equiv \kappa  + \delta \kappa$.
When $\delta \kappa \neq 0$, the lack of QND structure means that the optomechanical interactions can cause a net damping or antidamping of the system, even leading to instability. From Eq.~(\ref{SI_Eq_15}) we directly see that one requires $\delta \kappa > - \kappa/[\CC +1]$, to ensure the system is stable. For a large cooperativity, the stability condition is approximately $\delta \kappa > 0$, implying that the decay rate for cavity $2$ has to be larger as for cavity $1$ ($\kappa_2 > \kappa_1$).
The difference between the decay rates $\delta \kappa$ in Eq.~(\ref{SI_Eq_15}) is multiplied with the cooperativity $\CC$, and hence even small deviations from the symmetric $\kappa_1 = \kappa_2$ case can suppress the gain. Moreover, for $\CC \rightarrow \infty$ the resulting gain saturates at 
$( (\kappa_1 + \kappa_2)/( \kappa_2 - \kappa_1)) ^2$, and hence cannot be arbitrarily large as in the case for equal decay rates, cf. Fig.~2 of the main text. Even with this gain saturation at large $\CC$, the bandwidth $\lambda$ of the amplification remains finite in the limit of large gain and is even increased if $\kappa_1 \neq \kappa_2$. Based on Eq.~(\ref{SI_Eq_15}) and in the limit $\kappa \gg \gamma$ we can approximate the bandwidth as
\begin{align}\label{SI_Eq_16}
 \frac{\lambda}{\gamma} \approx  1 + \CC \frac{\delta\kappa}{\kappa + \delta \kappa}
              = 1 +  \CC  \left\{ \frac{\delta\kappa}{\kappa} + \mathcal O \left[ \frac{\delta\kappa^2}{\kappa^2} \right] \right\} ,
\end{align}
in the second step we performed an expansion for small deviations $\delta \kappa$. Assuming for example $\delta \kappa/\kappa = 1/\CC$ the resulting gain is reduced by a factor of $4$, while the bandwidth is increased by a factor of $2$.

With the reduction of the gain we also obtain a reduction in the noise. The symmetrized output noise becomes
\begin{align}\label{SI_Eq_17}
\bar S_{1,\rm out} [0] 	=& 
			\left[   \frac{ 2 \CC -1  + \frac{\delta \kappa}{\kappa} [\CC -1] } {   \frac{\delta \kappa}{\kappa}  [\CC  + 1] + 1 } \right]^2 
			 \hspace{-0.1cm} \left( \bar n_{d_{1}}^T + \frac{1}{2} \right) \hspace{-0.1cm}
				      +  \left( 1 + \frac{\delta \kappa }{\kappa} \right)
			\left[  \frac{ 2 \CC  } {   \frac{\delta \kappa}{\kappa}  [\CC  + 1] + 1  } \right]^2 
				  \hspace{-0.1cm} \left(\bar n_{d_{2}}^T  + \frac{1}{2} \right) \hspace{-0.1cm}
                                      + \left[  \frac{  2 \sqrt{\CC}   \left( 1 + \frac{\delta \kappa }{\kappa} \right) } 
						     { \frac{\delta \kappa}{\kappa}  [\CC  + 1] + 1}\right]^2 
			\hspace{-0.1cm}  \left( \bar n_b^T + \frac{1}{2} \right).
\end{align}
In the limit of a large cooperativity and small deviations $\delta\kappa$ we can estimate for the number of added noise quanta
\begin{align}\label{SI_Eq_18}
\bar n_{\rm add} =&  \ \bar n_{d_{2}}^T  + \frac{1}{2} + \frac{1}{\CC} \left( \bar n_{d_{2}}^T + \bar n_b^T   + 1 \right) 
					+ \frac{1}{2 \CC} \left( \bar n_{d_{2}}^T + 2 \bar n_b^T   + \frac{3}{2} \right)
					  \frac{\delta \kappa}{\kappa} + \mathcal O \left[\frac{\delta \kappa^2}{\kappa^2}, \frac{1}{\CC^2}\right] ,
\end{align} 
which clearly shows that for a large cooperativity we still reach the quantum limit. A small deviation in the decay rates has no significant effect on the added noise by the amplifier. From the expansion in Eq.~(\ref{SI_Eq_18}) we see that higher order contributions scale with $\delta \kappa / \CC$ and thus are suppressed for the large cooperativity we require to have large gain.

Additionally, it is worth noting that for unequal decay rates one also has the possibility of an effective normal-mode splitting in certain parameter regimes. 
In general, mode splitting in optomechanical systems appears in the strong coupling regime, where the interaction with the mechanical system dominates the dissipative interaction, i.e.,~ $G \gg \kappa,\gamma$ \cite{2013arXiv1303.0733ASI}. There a cavity mode hybridize with the mechanical mode and the resulting two normal modes perform coherent oscillations. The corresponding reflection spectra shows then two peaks at the frequencies of the normal modes. 
The situation is quite different here; for simplicity, consider the large $\gamma$ limit.
For a phase-conjugated coupling of the two modes alone (i.e.,~Eqs.~(3) of the main text without the $\pm \Gamma$ damping and antidamping terms), one finds analogous behavior (i.e.,~the phase conjugated coupling gives rise to oscillatory dynamics in the time domain). However, for symmetric $\kappa$ the additional damping and antidamping terms completely cancel this oscillatory tendency, as discussed in the main text.  Introducing a damping asymmetry spoils this cancellation, and hence oscillations (and mode splitting) are again possible.  For unequal $\kappa$, we find that mode splitting occurs when
\begin{align}\label{SI_Eq_19}
 G^2 > \frac{\kappa_1^2 \kappa_2^2 }{8 (\kappa_2^2 - \kappa_1^2 )} , 
\hspace{0.5cm} \mbox{for} \hspace{0.5cm} \kappa_2 > \kappa_1 \hspace{0.2cm} \mbox{and} \hspace{0.2cm}\gamma \ll \kappa_{1,2}.
\end{align}
In the regime of mode splitting, signal amplification still occurs, but now the gain exhibits two peaks as a function of frequency.  

\section{Unequal coupling strengths}
\noindent
While the main text focuses on  $G_1 = G_2$, it is also interesting to consider the case of unequal couplings.
Similar to the situation where the decay rates are unequal, the QND nature of the Hamiltonian is lost.  The result is that one can not longer achieve an arbitrarily large gain, as a large enough cooperativity can make the system unstable.

We start from the system Hamiltonian in Eq.~(2) in the main text and assume the same driving and dissipation terms as in the main text. With the former used description based on quantum Langevin equations and input-output theory we derive the photon number gain 
\begin{align}\label{SI_Eq_20}
S_{11}[\omega] =      \frac{\left(-\frac{4\omega^2}{\kappa^2} - 1 \right) \left(-i\frac{2\omega}{\gamma} + 1 \right)
                          -i\frac{2\omega}{\kappa} \CC_1 ( \RG^2 - 1) +  \CC_1 ( \RG^2 + 1) }
			    {\left(-i\frac{2\omega}{\kappa} + 1 \right)\left[
			      \left(-i\frac{2\omega}{\gamma} + 1 \right) \left(-i\frac{2\omega}{\kappa} + 1 \right) + \CC_1 (1 - \RG^2) \right]}
\hspace{0.3cm} \Rightarrow \hspace{0.3cm}
 \mathcal G_{1}[0] =  \left[ \frac{ \CC_1 ( \RG^2 + 1) - 1 }
			    {1 + \CC_1 (1 - \RG^2)} \right]^2, \ \RG \equiv \frac{G_2}{G_1},
\end{align}
which gives us the stability condition  $\RG < \sqrt{1+1/\CC_1}$, or equivalent $G_1^2 > G_2^2 - \kappa\gamma/4$ as in Ref.~\onlinecite{PhysRevLett.110.253601SI}.
Here, we defined the cooperativity as $\CC_1 = 4 G_1^2 / (\kappa \gamma)$ and for $\RG  \rightarrow 1 $, i.e.,~ $G_2 \rightarrow G_1$, we recover the result for equal coupling strengths. The consequences of unequal coupling strengths are similar to the case for unequal decay rates discussed in the last section. In the limit of a large cooperativity, the maximal gain saturates at $(G_2^2 + G_1^2)^2/(G_2^2 - G_1^2)^2 $. However, the
bandwidth $\lambda$ remains finite and is even increased compared to the case for equal strengths $G_1 = G_2$.
For $\kappa \gg \gamma$ we can approximate the frequency dependent gain resulting from Eq.~(\ref{SI_Eq_20}) and obtain for the corresponding bandwidth
\begin{align}\label{SI_Eq_21}
 \lambda \approx  \gamma \left(1 + \CC_1 [1 - \RG^2]\right)  = \gamma + \frac{4}{\kappa} \left[G_1^2 - G_2^2  \right] = \gamma + \Gamma_{\rm opt},
\end{align}
i.e., the bandwidth is just the mechanical damping plus the optical damping $\Gamma_{\rm opt}$.
Thus, if we choose for example $1 - \RG^2 = 1 / \sqrt{\CC_1}$ and assume a large cooperativity, the resulting gain is $\mathcal G_1 \approx  4 \CC_1$, while the bandwidth is increased to
$\lambda \approx \gamma \sqrt{\CC_1}$. Nevertheless, we have to care about possible mode splitting as in the case for unequal decay rates.
Note, that when $G_1 \neq G_2$, we have basically the same system studied in Ref.~\onlinecite{PhysRevLett.110.253601SI}, and the arising normal modes here are the same as those discussed in that work. The eigenvalues of the susceptibility matrix, i.e.,~Eq.~(\ref{SI_Eq_2}), are now given by:
\begin{align}\label{SI_Eq_22}
\epsilon_{1}   = \frac{2}{\kappa},
\hspace{0.5cm}
 \epsilon_{2,3} = \frac{1}{ 1 + \CC_1 [1- \RG^2] } \left[ \frac{1}{\gamma} + \frac{1}{\kappa} 
			\mp \sqrt{\left(\frac{1}{\kappa}-\frac{1}{\gamma}\right)^2 -\frac{4 \CC_1}{\kappa \gamma} [1 - \RG^2] } \right],
\end{align}
and if $\epsilon_{2,3}$ have a non-zero imaginary part, two additional peaks show up in the corresponding reflection spectra. 
For $\RG = 1$ we recover the real eigenvalues of Eq.~(\ref{SI_Eq_3}), which are independent of the coupling strength $G = G_1 = G_2$. Hence, in the symmetric parameter regime (i.e.,~for equal coupling strengths and equal decay rates) we find no mode splitting at all. 
Otherwise, to avoid the effect of mode splitting for $G_1 \neq G_2$ we have to consider the condition $G_1^2 - G_2^2 < (\kappa - \gamma)^2/4 $. 
For $\gamma \ll \kappa $ we can approximate this condition as $\Gamma_{\rm opt} < \kappa$ and hence the maximal obtained bandwidth is $\lambda_{\rm max} \simeq \kappa $, cf.~Eq. \ref{SI_Eq_21}. Thus, for the upper example, i.e., $\Gamma_{\rm opt} = \gamma \sqrt{\CC_1}$, the maximal amplitude-gain bandwidth product is $2\kappa^2/\gamma$.

The zero-frequency noise of the output signal reads
\begin{align} \label{SI_Eq_23}
\bar S_{1,\rm{out}} =  
                   \left[ \frac{ \CC_1 ( \RG^2 + 1) - 1 }{1 + \CC_1 (1 - \RG^2)} \right]^2 
                   \hspace{-0.1cm} \left( \bar n_{d_{1}}^T + \frac{1}{2} \right) \hspace{-0.1cm} 
		  +\left[ \frac{ 2 \CC_1 \ \RG }{1 + \CC_1 (1 - \RG^2)} \right]^2
		   \hspace{-0.1cm} \left(\bar n_{d_{2}}^T  + \frac{1}{2} \right) \hspace{-0.1cm}
		  + \left[ \frac{ 2 i \sqrt{\CC_1}  }{1 + \CC_1 (1 - \RG^2)} \right]^2
		 \hspace{-0.1cm}  \left( \bar n_b^T + \frac{1}{2} \right),
\end{align}
which is reduced as the gain. Finally, in the limit of large gain and for small deviations between the coupling strengths the added noise quanta yield
\begin{align}\label{SI_Eq_24}
 \bar n_{\rm add} = \bar n_{d_{2}}^T  + \frac{1}{2} + \frac{1}{\CC_1} \left(n_{d_{2}}^T + \bar n_b^T + 1 \right)
                     + \frac{1}{\CC_1} \left(\bar n_{d_{2}}^T   + 2 \bar n_b^T + \frac{3}{2} \right) [1 - \RG ] 
                      + \mathcal O\left[(1 - \RG)^2, \frac{1}{\CC_1^2} \right],
\end{align}
which clearly shows that we still can reach the quantum limit.


\section{Counter-rotating terms} 
\noindent
We now consider the effects of the nonresonant, counter-rotating terms on our dissipative parametric amplification scheme; these are described by:
\begin{align}\label{SI_Eq_25}
 \hH_{\rm CR}
 =&  G \left\{  \left(\hat d_1^{\dag} \hat b^{\dag}+ \hat d_2^{\PD}  \hat b^{\dag}\right) e^{i2\wM t} 
			          + \left(\hat d_1^{\PD}  \hat b       + \hat d_2^{\dag} \hat b       \right) e^{-i2\wM t} \right\}.
\end{align}
This Hamiltonian describes Stokes (anti-Stokes) processes for the photons in cavity $2 (1)$ which are neglected in RWA. 
The full non-RWA theory is exactly solvable by working in an interaction picture where the total Hamiltonian is time-independent.

Taking the counter-rotating terms into account, we perform our standard calculation based on input-output theory to derive the photon number gain on resonance 
\begin{align}\label{SI_Eq_26}
\sqrt{\mathcal G_1 [0]}  =&
		\frac{ 2 \CC - 1 +
		 \mathcal C    \frac{\delta \kappa}{\kappa}  \left[1
		 +  \frac{1}{(-4i \frac{\wM}{\kappa + \delta \kappa} + 1) (-4i \frac{\wM}{\kappa} +1)}\right]
		 -  \frac{i}{4}  \frac{\gamma}{\wM}  }
		{\mathcal C    \frac{\delta \kappa}{\kappa}  \left[1
		 -  \frac{1}{(-4i \frac{\wM}{\kappa + \delta \kappa} + 1) (-4i \frac{\wM}{\kappa} +1)}\right]
		+ 1 +  \frac{i}{4}  \frac{\gamma}{\wM} }.
\end{align}
Using this, we can estimate that if the decay rates are unequal the counter-rotating terms have more influence than in the ideal, symmetric case. Similar behavior is found
for unequal coupling strengths.
But for $\delta \kappa =0 $ the terms in the square brackets in Eq.~(\ref{SI_Eq_26}) disappear and we obtain 
\begin{align}\label{SI_Eq_27}
\mathcal G_1 [0]  =&
		\frac{(2\mathcal C   - 1)^2  +  \frac{\gamma^2}{16}  \frac{1}{\wM^2} }
		{1 +  \frac{\gamma^2}{16}  \frac{1}{\wM^2} }
		= (2\mathcal C   - 1)^2 + 4 \mathcal C ( \CC -1 ) \ \sum_{n = 1}^{\infty} \ (-1)^{n} \left( \frac{\gamma/4 }{\wM} \right)^{2n} ,
\end{align}
therewith we have a scaling with $\gamma/\wM$ and the cooperativity $\mathcal C$ remains a linear factor and does not enhance higher order terms. It is interesting to note that these terms do not depend on the decay rate $\kappa$; $\wM \gg \gamma$ is the only relevant condition to keep the unwanted side-band processes off-resonant. We stress that this conclusion only holds for the gain, and for the ideal symmetric case $G_1 = G_2$ and $\kappa_1 = \kappa_2$.
For the number of added noise quanta the RWA and non-RWA results require $\wM \gg \kappa$ to coincide even in the symmetric case, as shown in Fig. \ref{Img_3}, which compares the full theory against the approximate RWA results. The added noise in non-RWA contains as well noise incident on cavity $1$ due to next sideband contributions.
Moreover, deviations from the ideal QND situation require $\wM \gg \kappa$ if one wishes the RWA to be valid.

\begin{figure}[t]
  \begin{center}
    \includegraphics[width=1.0\textwidth]{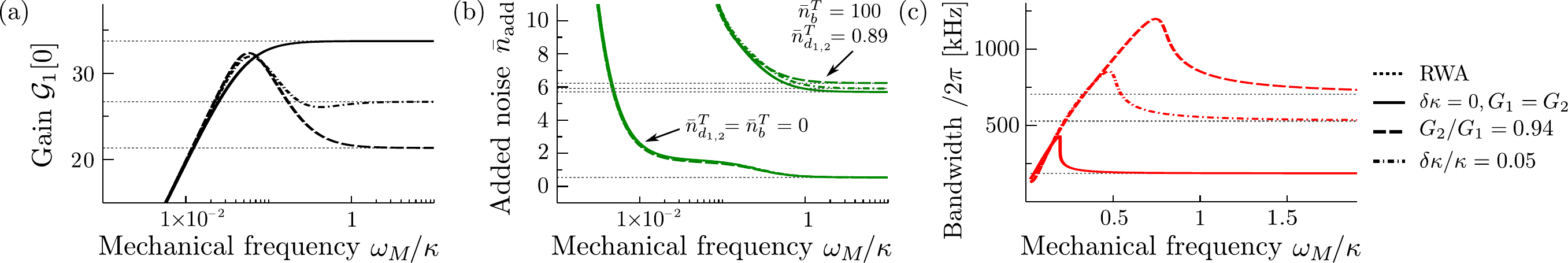}
  \end{center}
 	\caption{Influence of counter-rotating terms on the relevant system quantities gain, added noise quanta and bandwidth.
		The solid lines show the ideal symmetric case, while the dashed-dotted (dashed) line includes the effect of
		 asymmetric cavity damping (coupling strengths), for parameters see legend. The horizontal dotted lines correspond to the respective RWA result.
		We choose $\kappa / (2 \pi)  = 1 \textrm{ MHz}$, which includes internal loss $\kappa^{\rm int}/( 2 \pi)  = 10 \textrm{kHz}$, and a
		cooperativity of $\CC = 25$. We assume the mechanical mode to be optically cooled, leading to the effective damping
		 $\gamma / (2 \pi) = 200 \textrm{ kHz}$.
		(a) Photon number gain as a function of $\wM/\kappa$. 
		(b) Added noise quanta versus $\wM$. The lower lines correspond to zero temperature baths (without intrinsic losses); 
		and the lines above to the the case where we include thermal noise on both the cavities and the mechanics, with $\bar n^T_{b,d_2}$ 
		as denoted in the graph.
		(c) Bandwidth versus the mechanical frequency, which is defined as the full-width at half-maximum of $\mathcal{G}_1[\omega]$. 
		Note, in the asymmetric cases the resulting bandwidth is significantly increased, but by the price
		 of a reduction in photon number gain, cf. graph (a).
		\label{Img_3}}
\end{figure}



\section{Superconducting circuit realization of dissipative amplification}
\noindent
As mentioned in the main text (and sketched in Fig.~1(b)), the dissipative amplification scheme we describe could also be readily implemented in superconducting
circuits utilizing Josephson junctions.  The basic idea is to use the nonlinearity of a Josephson junction (essentially a nonlinear inductance) to realize the kind of
parametric interactions required in our basic interaction Hamiltonian (Eq.~(2) of the main text).  A key advantage over the optomechanical realization is that all three bosonic
modes (signal, idler and auxiliary ``$b$" mode) will be microwave modes.  As such, having the auxiliary mode damping rate (and hence the amplification bandwidth)
 be in the range of several MHz is not difficult to achieve.

\begin{figure*}[ht]  
 \includegraphics[width=1.0\textwidth]{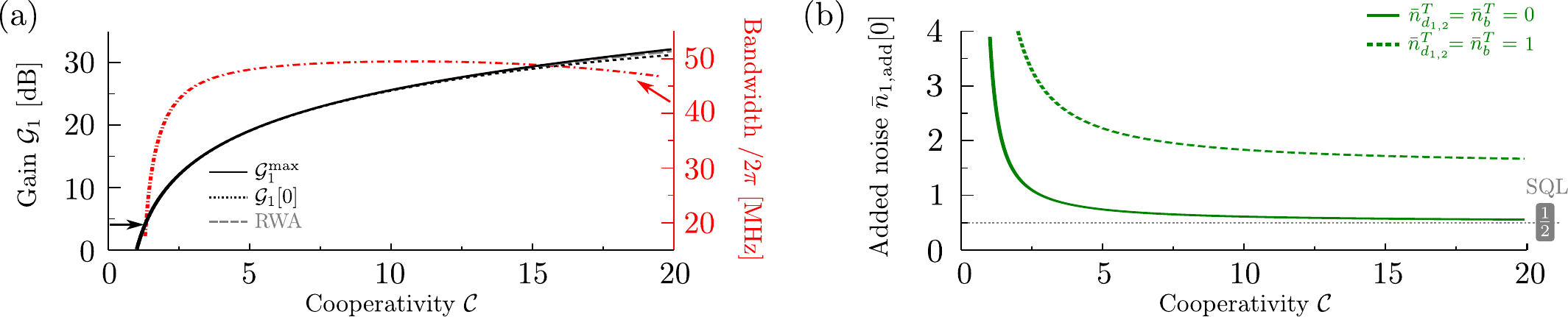}
	\caption{(a) Black curves:  photon number gain versus cooperativity $\mathcal{C}$ for parameters corresponding to a superconducting circuit realization of the dissipative amplification scheme \cite{Abdo2013b}.  We take $\wM / (2 \pi) = 1 \textrm{ GHz}$, $\omega_1 = \omega_2 = 10 \textrm{ GHz}$
	and $\kappa / (2 \pi)  = \gamma / (2 \pi) = 100 \textrm{ MHz}$.   
	The dotted line depicts the zero frequency gain $\mathcal G_1[0]$; the grey dashed line shows
	the RWA result and the solid line corresponds to the actual maximum $\mathcal G_1^{\rm max}$.  
	Red dashed-dotted curve:  bandwidth (defined as the full-width at half-maximum of $\mathcal{G}_1[\omega]$)
	 versus $\mathcal{C}$, same parameters as solid curve. 
	(b)  Amplifier added noise $\bar{n}_{\rm add}$ versus $\mathcal{C}$.  
	Solid curve: vacuum noise only.  Dashed curve:  
	thermal noise with $\bar n^T_{b,d_{1,2}} = 1$ as denoted in the graph. 
	The shown results include the relevant counter-rotating terms, i.e.,~ the leading sidebands associated with all counter-rotating terms in Eq. \ref{eq:HSCCR}, which lead to a shift of the gain maximum $\mathcal G_1^{\rm max}$ from the zero frequency result $\mathcal G_1[0]$, as well as a decrease of the bandwidth for higher values of $\mathcal C$. 
		\label{Img_4}}
\end{figure*}

While several routes are possible, we imagine a  implementation using 
the Josephson parametric converter (JPC), as pioneered in Ref.~\onlinecite{Bergeal2010aSI}.  The JPC is a symmetric circuit involving four Josephson junctions in a ring geometry
that realizes an interaction involving three bosonic 
modes of the form
\begin{equation}
	\hH_{\rm JPC} = \Lambda \left( \hat{a}_1^{\PD} + \hat{a}_1^\dagger \right) \left( \hat{a}_2^{\PD} 
			+ \hat{a}^\dagger_2 \right) \left( \hat{a}_3^{\PD} + \hat{a}^\dagger_3 \right),
\end{equation}
where $\hat{a}_j$ ($j=1,2,3$) are the annihilation operators for each mode.  We now imagine a circuit having two JPCs with one shared mode:
\begin{align}\label{Eq_13}
 \hH_{\rm SC} =& \sum_{j \in{1,2}} \left( \omega_{j,  P} \ \hat a_{j}^{\dag} \hat a_{j}^{\PD} 
					  +  \ \omega_j \ \hat d_j^{\dag} \hat d_j^{\PD} \right)
		+ \omega_{ M} \ \hat b^{\dag} \hat b
		+ g_1  \left(  \hat d_1^{\PD} + \hat d_1^{\dag} \right) \left(\hat b + \hat b^{\dag} \right) \left( \hat a_1^{\PD} + \hat a_1^{\dag}\right)
                + g_2  \left(  \hat d_2^{\PD} + \hat d_2^{\dag} \right) \left(\hat b + \hat b^{\dag} \right) \left( \hat a_2^{\PD} + \hat a_2^{\dag}\right).
\end{align}
Here, the lowering operators of the five modes are $\hat{a}_{1,2}, \hat{d}_{1,2}, \hat{b}$, and $g_{1,2}$ are proportional to the Josephson energies of each JPC.  The operator $\hat{b}$ (with resonance frequency $\omega_{M}$) describes the shared common mode between the two JPCs; it will play an analogous role to the $\hat b$ mode in the main text (the mechanical mode in an optomechanical realization), and will be used to mediate a dissipative interaction.  The modes $\hat{d}_{1,2}$ (resonance frequencies $\omega_{j}$, $j=1,2$) will play the role of signal and idler modes, also in complete analogy to the main text.  Finally, the modes $\hat{a}_{j}$ 
(resonance frequencies $\omega_{j, P}$) will be pump modes:  by strongly driving these at the appropriate frequency, we can realize parametric interactions between $\hat{d}_{1,2}$ and $\hat{b}$ given in the central interaction Hamiltonian of the main text, Eq.~(2).  We take the pump mode frequencies to be $\omega_{1,  P} = \omega_1 - \omega _{ M}$ and $\omega_{2,  P} = \omega_2 + \omega_{M}$, and assume that each pump mode is strongly driven on resonance.  The strong drive lets us linearize the above Hamiltonian by replacing
the pump-mode operators $\hat{a}_j$ by their average value $\bar{a}_j$.  Working in an interaction picture with respect to the free mode Hamiltonians, and taking $g_1 = g_2 \equiv g$ as well as $\bar{a}_1 = \bar{a}_2 \equiv \bar{a}$ for simplicity, we find:
\begin{align}
	\label{eq:HSCMain}
\hH_{\rm SC} =& \ G  \left( \hat d_1^{\PD} \hat b^{\dag}+\hat d_1^{\dag} \hat b+\hat d_2^{\PD}\hat b +\hat d_2^{\dag} \hat b^{\dag}  \right)
			+   \hH_{\rm SC, CR} ,
 \\   
		\label{eq:HSCCR}
\hH_{\rm SC, CR}  =&  \ G \left\{ \left(\hat d_1^{\dag} \hat b^{\dag}+ \hat d_2^{\PD}  \hat b^{\dag}\right) e^{i2 \omega _{ M} t} 
				  + \hat d_1^{\dag} \hat b        e^{ i 2 \omega_{1,  P} t}
				  + \hat d_2^{\dag} \hat b^{\dag} e^{ i 2 \omega_{2,  P} t}
				  + \hat d_1^{\dag} \hat b^{\dag} e^{ i 2 \omega_1 t} 
				  + \hat d_2^{\dag} \hat b        e^{ i 2 \omega_2 t} + h.c. \right\},
\end{align}
where $G = g \bar{a}$ is the pump-enhanced parametric interaction strength.  The terms in $\hH_{\rm SC, CR}$ describe non-resonant interactions and will have minimal impact if the associated frequencies are much larger than the damping rates $\kappa$ of the modes $\hat{d}_j, \hat{b}$.  In this limit, we see that the dual JPC system achieves the same Hamiltonian given in 
Eq.~(2) of the main text, i.e.,~the basic interaction Hamiltonian which gives rise to dissipative amplification.

While the interaction Hamiltonian in the RWA is identical to that derived in the main text for the optomechanical realization, the non-RWA terms in Eq.~(\ref{eq:HSCCR}) are not identical
to the non-RWA terms arising in the optomechanical realization, c.f.~Eq.~(\ref{SI_Eq_25}).  In particular, the terms oscillating at twice the pump-mode frequencies $\omega_{j,  P}$ do not appear in the optomechanical setup.  We find that the impact of these terms on the perfect dissipative amplification physics arising from the Hamiltonian in Eq.(2) is more severe than other non-RWA terms.  
In practice, this means that the maximum achievable photon number gain will be more limited by non-RWA terms in the JPC realization than the optomechanical realization.

Nevertheless, this circuit realization of our dissipative amplification scheme could allow one to surpass the current state of the art in Josephson junction paramps \cite{Abdo2013b, Mutus2013b}. As demonstrated by Mutus et al.~\cite{Mutus2013b}, with a lumped-element Josephson-junction parametric amplifier, it is possible to achieve a bandwidth of $\sim 50\textrm{ MHz}$ for a gain of $20 \textrm{ dB}$, by using a highly damped microwave cavity ($Q \sim 10,\kappa \sim 1 \textrm{GHz}$). Our proposed setup can match this performance easily starting with
\emph{a much higher $Q$ microwave cavity}.
As shown in Fig.~\ref{Img_4},  
with $\mathcal C \sim 6$ we obtain a gain of $20 \textrm{ dB}$, while the bandwidth is approximately $\kappa/2$, for a bandwidth of $50\textrm{ MHz}$ and $\kappa = 100 \textrm{ MHz}$. 
The fact that our scheme allows such performance with a higher $Q$ ($Q_1 \sim 100$ versus $Q \sim 10$ in Ref.~\onlinecite{Mutus2013b}) has may significant practical advantages.  
For example, using low-$Q$ resonances means that very large pump powers are required for amplification, which in turn could lead to unwanted effects, e.g.,  the resulting large cavity amplitudes can lead to additional dissipative processes, as well as higher nonlinearities in the Josephson potential can become important. In contrast, the dissipative amplification scheme shows a good performance for higher $Q$ values, without damaging the bandwidth of the amplified signal.  The results shown in Fig.~\ref{Img_4}, can be further improved with small changes in parameters.  For example, by now choosing 
$\wM = 2.5 \textrm{ GHz}$, $\kappa = 250 \textrm{ MHz}$, $\gamma = 125 \textrm{ MHz}$, we obtain a bandwidth of $90\textrm{ MHz}$ for a gain of $20 \textrm{ dB}$.  This results in a amplitude-gain bandwidth product that is a factor of two better than in Ref.~\onlinecite{Mutus2013b}, while still having a signal cavity with $Q\sim 40$.
Another way of increasing the bandwidth is possible by choosing asymmetric coupling strengths, e.g. taking the same parameters as in Fig.~\ref{Img_4} but with $G_1/G_2 = 0.94$ results in a bandwidth of $120$~MHz for $21$~dB gain.

Perhaps even more advantageous than the ability to match existing amplifier performance with large cavity $Q$ is the ability to achieve much higher gains without any consequent decrease in bandwidth.  For example, as shown in Fig.~\ref{Img_4}, a small increase in $\mathcal{C}$ allows to increase the gain from $20 \textrm{ dB}$ to $30 \textrm{ dB}$, while maintaining a bandwidth of 
$50\textrm{ MHz}$.  Such higher gains are desirable as they greatly reduce the requirements on the following amplifier and any associated insertion losses; this is particularly important in state-of-the-art experiments requiring near quantum-limited performance.  Consider for example applications in continuous-measurement based quantum feedback, as implemented in recent experiments \cite{Vijay2012b,Hatridge2013b}.
Such protocols have a measurement efficiency $\eta = 1/ 2 \bar{n}_{\rm add}$, where $\bar{n}_{\rm add}$ is the total added noise including the effects of following amplifiers.  A standard HEMT following amplifier adds anywhere from $20$ to $50$ noise quanta (when one includes typical insertion losses).  A paramp gain of $20 \textrm{ dB}$ means that this noise, referred back to the signal, yields $50 / 100 = 0.5$ noise quanta on top of the paramp noise contribution.  Thus, even if the paramp is quantum limited, the total added noise is twice the quantum limit value, implying $\eta \sim 0.5$.  In contrast, using a gain of $30 \textrm{ dB}$  (as is achieved by our scheme with no bandwidth degradation) improves this efficiency to $\eta = 0.95$.  Such an improvement in measurement efficiency can dramatically increase the power of measurement-based feedback protocols (see, e.g., Refs.~\onlinecite{Vijay2012b,Hatridge2013b}).

Finally, note that the optimal high-bandwidth performance shown in Fig.~\ref{Img_4} could in principle also be reached in microwave-cavity optomechanical systems, if one now made use of a high-frequency, low-$Q$ mechanical resonator (i.e.,~ with a damping rate $\gamma \sim \kappa \sim 100 \textrm{ MHz}$).  Resonators of this sort were recently used in pioneering experiments by
O'Connell et al. \cite{Cleland2010b}, where they were coupled to microwave-frequency superconducting qubits.  Similar approaches could be used to achieve to couple them to microwave-frequency cavities, leading to a possible realization of our scheme.  Note that in such a system, the fact that mechanical and cavity frequencies are comparable implies that the same additional counter-rotating terms found in the JPC setup will be relevant, i.e.,~Eq. \ref{eq:HSCCR}.
\\
\\

\end{widetext}

\end{document}